\PassOptionsToPackage{unicode}{hyperref}
\PassOptionsToPackage{hyphens}{url}
\documentclass[
  11pt,
  a4paper]{article}
\usepackage{xcolor}
\usepackage[margin=1in]{geometry}
\usepackage{amsmath,amssymb}
\usepackage{iftex}
\ifPDFTeX
  \usepackage[T1]{fontenc}
  \usepackage[utf8]{inputenc}
  \usepackage{textcomp} 
\else 
  \usepackage{unicode-math} 
  \defaultfontfeatures{Scale=MatchLowercase}
  \defaultfontfeatures[\rmfamily]{Ligatures=TeX,Scale=1}
\fi
\usepackage{lmodern}
\ifPDFTeX\else
\fi
\IfFileExists{upquote.sty}{\usepackage{upquote}}{}
\IfFileExists{microtype.sty}{
  \usepackage[]{microtype}
  \UseMicrotypeSet[protrusion]{basicmath} 
}{}
\makeatletter
\@ifundefined{KOMAClassName}{
  \IfFileExists{parskip.sty}{%
    \usepackage{parskip}
  }{
    \setlength{\parindent}{0pt}
    \setlength{\parskip}{6pt plus 2pt minus 1pt}}
}{
  \KOMAoptions{parskip=half}}
\makeatother
\usepackage{longtable,booktabs,array}
\usepackage{calc} 
\usepackage{etoolbox}
\makeatletter
\patchcmd\longtable{\par}{\if@noskipsec\mbox{}\fi\par}{}{}
\makeatother
\IfFileExists{footnotehyper.sty}{\usepackage{footnotehyper}}{\usepackage{footnote}}
\makesavenoteenv{longtable}
\NewDocumentCommand\citeproctext{}{}
\NewDocumentCommand\citeproc{mm}{%
  \begingroup\def\citeproctext{#2}\cite{#1}\endgroup}
\makeatletter
 \let\@cite@ofmt\@firstofone
 \def\@biblabel#1{}
 \def\@cite#1#2{{#1\if@tempswa , #2\fi}}
\makeatother
\newlength{\cslhangindent}
\newlength{\csllabelwidth}
\newenvironment{CSLReferences}[2] 
 {\begin{list}{}{%
  \setlength{\itemindent}{0pt}
  \setlength{\leftmargin}{0pt}
  \setlength{\parsep}{0pt}
  \ifodd #1
   \setlength{\leftmargin}{\cslhangindent}
   \setlength{\itemindent}{-1\cslhangindent}
  \fi
  \setlength{\itemsep}{#2\baselineskip}}}
 {\end{list}}
\usepackage{calc}

\providecommand{\tightlist}{%
  \setlength{\itemsep}{0pt}\setlength{\parskip}{0pt}}
\usepackage{bookmark}
\IfFileExists{xurl.sty}{\usepackage{xurl}}{} 
\hypersetup{
  pdftitle={Separating Engineering Reasoning from DEXPI Serialization in LLM-Based Greenfield Surface-Process Design: A Three-Case Study for Underground Gas Storage},
  pdfauthor={Qingchuan Zhu (1, corresponding author); Shuyue Tong (1); Pengju Ren (2)},
  hidelinks,
  pdfcreator={LaTeX via pandoc}}

\title{Separating Engineering Reasoning from DEXPI Serialization in
LLM-Based Greenfield Surface-Process Design: A Three-Case Study for
Underground Gas Storage}
\author{Qingchuan Zhu (1, corresponding author) \and Shuyue Tong
(1) \and Pengju Ren (2)}
\date{}

\begin{document}
\maketitle

\begin{enumerate}
\def\labelenumi{\arabic{enumi}.}
\tightlist
\item
  Sinopec Petroleum Engineering Zhongyuan Co., Ltd., No.~33 Zhengguang
  North Street, Zhengdong New District, Zhengzhou, Henan, China
\item
  Xi'an Jiaotong University, No.~28 Xianning West Road, Xi'an, Shaanxi
  710049, P.R. China
\end{enumerate}

Corresponding author: Qingchuan Zhu Email: zhuqingchuan@foxmail.com
ORCID: 0000-0003-1816-070X

\begin{abstract}
Large language models can produce engineering descriptions and structured process representations, but standards-level serialization can substantially increase the generation burden. This diagnostic study examines whether separating engineering reasoning from Data Exchange in the Process Industry (DEXPI) serialization changes where representation and engineering failures occur in constrained greenfield surface-process design for underground gas storage. We compare Direct DEXPI generation with generation of a lightweight Engineering Intermediate Representation (IR) on three cases: single-pressure injection, withdrawal and export, and dual-pressure injection. All six conditions use one fixed model snapshot, qwen3.8-max-0902, with one completed hosted generation per condition. Direct outputs are XSD-valid in 2 of 3 cases, while all 3 Engineering IR outputs are structurally valid under a minimal validator. Direct prompt inputs contain approximately 121.8k–121.9k tokens, compared with 617–699 tokens for the Engineering IR prompts. Engineering feasibility does not uniformly favor the IR: one IR output is rejected for an explicit cooling-state contradiction. The cases also reveal two distinct Direct DEXPI failure modes: engineering inconsistency and standards-level serialization failure. The observed comparison shows that, in these evaluated method bundles, deferring DEXPI serialization substantially reduces representation burden and helps isolate serialization failure, but reducing representation burden alone does not eliminate engineering inconsistencies.
\end{abstract}

\section{1. Introduction}\label{introduction}

Greenfield process design requires several kinds of correctness at the
same time. A proposed process must represent an intended sequence of
operations, preserve material-flow and state relationships, and satisfy
explicit boundary conditions. When the design is also required to be
delivered in a standards-level data format, the model must solve
engineering reasoning and serialization in the same generation.

This study examines a narrow version of that problem for surface
processes serving underground gas storage (UGS). The cases are
deliberately constrained: pipeline-quality natural gas is injected into
storage wells, withdrawn for export, or delivered simultaneously to two
injection destinations with different pressure requirements. The design
basis supplies capacity, pressure, temperature, and boundary
constraints, while the model chooses the major operations and
intermediate states.

Two output methods are compared. In the Direct condition, the model is
asked to return a complete Data Exchange in the Process Industry (DEXPI)
2.0 Process XML document while using the frozen official reference
artifacts. In the Engineering IR condition, the model is asked to return
a lightweight JSON representation containing nodes and streams with
pressures, temperatures, and flows. The latter condition postpones
standards-level serialization.

The research question is:

How do representation burden and observable failure modes differ when
engineering design generation and standards-level DEXPI serialization
are performed jointly versus when DEXPI serialization is deferred?

The purpose is diagnostic: to observe how the two representation
strategies expose different engineering and serialization failure modes.

\section{2. Background and Related
Work}\label{background-and-related-work}

Recent chemical-process research has examined large language models and
LLM agents for process simulation, flowsheet autocorrection, and
automated flowsheet generation. Liang et al.~connect an LLM agent to a
rigorous process simulator for natural-language flowsheet analysis and
synthesis \citeproc{ref-liang2026large}{Liang et al. 2026}; Schulze
Balhorn et al.~use SFILES 2.0 to represent PFD/P\&ID flowsheets as
strings for autocorrection with language models
\citeproc{ref-schulze_balhorn2024autocorrection}{Balhorn et al.
2024}; and Sketch2Simulation constructs a graph-based intermediate
representation before code generation and simulator execution
\citeproc{ref-bahamdan2026sketch2simulation}{Bahamdan et al. 2026}.
These studies illustrate growing use of structured process-engineering
representations that can be parsed and checked rather than treated only
as free-form text. The use of an intermediate representation is
therefore not claimed as novel here.

More directly, Gowaikar et al.~describe an agentic P\&ID-generation
workflow in which natural-language inputs are organized into an
intermediate domain-specific language and then translated
deterministically into DEXPI/Proteus XML
\citeproc{ref-gowaikar2024agentic}{Gowaikar et al. 2024}. Laub et
al.~convert natural-language process descriptions into machine-readable
flowsheet graphs and use SFILES as an intermediate representation that
can be stored or exchanged in standardized formats including DEXPI
Process \citeproc{ref-laub2026texttoflowsheet}{Laub et al. 2026}.
These studies establish prior use of both DEXPI-targeted generation
workflows and intermediate representations in process-engineering
automation.

DEXPI provides a standardized, machine-readable representation of
process-engineering information
\citeproc{ref-dexpi_specification_2_0_0}{DEXPI Initiative and DEXPI
e.V. 2025}. If an LLM can generate valid DEXPI directly, the result
can in principle be exchanged with DEXPI-compatible engineering tools,
checked programmatically, transformed into other engineering
representations, or used as input to downstream automation without first
being manually re-entered or restructured. This makes Direct DEXPI
generation attractive as a potential path from language-model reasoning
to interoperable engineering data. At the same time, standards-level
representation adds more than a generic output format: engineering
objects and their relationships must be expressed within the conventions
of the information model and its XML serialization.

Structured output constraints can make model responses easier to parse
and inspect, but structure alone does not eliminate errors in
engineering reasoning. Direct DEXPI generation asks the language model
to handle engineering semantics, the standards-level information model,
and XML serialization within the same generation. Against this
background, the present study performs a matched diagnostic comparison
of two method bundles under the same fixed model snapshot and on the
same three cases: Direct DEXPI generation versus a lightweight
Engineering Intermediate Representation (IR). It separately records
representation validity, engineering feasibility, and method-specific
completion while asking how failure modes differ when DEXPI
serialization is performed jointly with engineering design generation or
deferred to a later step.

\section{3. Study Design}\label{study-design}

\subsection{3.1 Generation methods}\label{generation-methods}

Each design basis is supplied to both methods. The Direct DEXPI prompt
asks for one complete DEXPI 2.0 Process XML document and includes the
frozen official DEXPI 2.0.0 schema and model files as reference context.
The Engineering IR prompt asks for one JSON object with exactly five
top-level fields: \texttt{case\_name}, \texttt{scope},
\texttt{material}, \texttt{nodes}, and \texttt{streams}. Each stream
records flow, pressure, and temperature.

The exact prompt templates used in the formal runs are preserved as
\path{runtime/prompts/formal_direct_dexpi_v0.1.txt} and
\path{runtime/prompts/formal_engineering_ir_v0.1.txt}.

The two conditions are treated as method bundles rather than as an
isolated XML-versus-JSON comparison. The Direct condition includes the
large DEXPI reference context and a strict XML-only output contract,
whereas the IR condition uses a much smaller schema and a different
output contract. The comparison therefore examines practical differences
between the two method bundles rather than isolating a single
independent variable.

\subsection{3.2 Generation settings and
protocol}\label{generation-settings-and-protocol}

The formal study uses one fixed model snapshot:

\begin{itemize}
\tightlist
\item
  model: qwen3.8-max-0902
\item
  reasoning effort: none
\item
  temperature: 0.7
\item
  top\_p: 0.8
\item
  top\_k: 20
\item
  repetition penalty: 1.0
\item
  seed: 1234
\item
  maximum completion tokens: 16,384
\item
  n: 1
\item
  preserve\_thinking: false
\item
  stream: false
\end{itemize}

Hosted inference was performed through Alibaba Cloud DashScope using its
OpenAI-compatible API. Formal runs used Python 3.10.11 and OpenAI SDK
2.54.0.

There is one completed hosted generation per case-method condition. No
result-dependent retry, repair, or regeneration was applied. The seed is
a reproducibility control, not a guarantee of bitwise deterministic
hosted inference.

\subsection{3.3 Evidence preservation}\label{evidence-preservation}

The reproducibility package preserves the generated XML or IR JSON, run
metadata, and the corresponding validation record for all six
conditions. The exact generated files are treated as raw evidence. They
are not normalized or rewritten for presentation. File-level SHA-256
hashes for the frozen formal artifacts are provided in
\texttt{FORMAL\_ARTIFACT\_SHA256SUMS.txt}.

\section{4. Test Cases}\label{test-cases}

Across all three cases, the model independently determines the major
operations, their sequence, and the intermediate process states required
to satisfy the specified boundary conditions. Unless otherwise stated,
no particular process configuration is prescribed.

{\def\LTcaptype{none} 
\begin{longtable}[]{@{}
  >{\raggedright\arraybackslash}p{(\linewidth - 6\tabcolsep) * \real{0.2500}}
  >{\raggedright\arraybackslash}p{(\linewidth - 6\tabcolsep) * \real{0.2500}}
  >{\raggedright\arraybackslash}p{(\linewidth - 6\tabcolsep) * \real{0.2500}}
  >{\raggedright\arraybackslash}p{(\linewidth - 6\tabcolsep) * \real{0.2500}}@{}}
\toprule\noalign{}
\begin{minipage}[b]{\linewidth}\raggedright
Case
\end{minipage} & \begin{minipage}[b]{\linewidth}\raggedright
Feed
\end{minipage} & \begin{minipage}[b]{\linewidth}\raggedright
Capacity / split
\end{minipage} & \begin{minipage}[b]{\linewidth}\raggedright
Required outlet
\end{minipage} \\
\midrule\noalign{}
\endhead
\bottomrule\noalign{}
\endlastfoot
Case 1: single-pressure injection & 8.0--10.0 MPa, approximately 2 °C &
700,000 standard m³/day & Injection wells at up to 35 MPa; temperature
≤60 °C \\
Case 2: withdrawal and export & 12.0--14.0 MPa, approximately 25 °C &
450,000 standard m³/day & Export at 6.5--7.5 MPa; temperature ≤45 °C \\
Case 3: dual-pressure injection & 8.0--9.0 MPa, approximately 15 °C &
900,000 standard m³/day total: 300,000 to Group A and 600,000 to Group B
& Group A at 20 MPa and Group B at 30 MPa; both temperatures ≤55 °C \\
\end{longtable}
}

For Case 2, no additional gas-quality treatment is specified.

\section{5. Evaluation}\label{evaluation}

For Direct DEXPI, Representation Validity is PASS when the XML is
parseable and valid against the frozen official XSD using
evaluation/validate\_xsd.py. It is FAIL when parsing or XSD validation
fails. XSD validity is the Direct representation-validity gate and does
not establish full DEXPI semantic conformance.

For Engineering IR, Representation Validity is PASS when the JSON is
parseable and valid under the minimal IR structural validator retained
in runtime/run\_ir\_once.py. IR PASS means only structurally valid under
that validator; it is not a physics or engineering-validity claim.

Engineering Feasibility uses three labels:

\begin{itemize}
\tightlist
\item
  ACCEPT
\item
  WARN
\item
  REJECT
\end{itemize}

Within Engineering Feasibility, WARN denotes an otherwise acceptable
engineering result with a non-fatal concern.

The engineering review considers process topology, flow and state
continuity, operation/state consistency, and the stated boundary
conditions. It checks whether streams are connected, whether required
flows are represented, whether pressure and temperature changes agree
with named operations, and whether final boundary states satisfy the
case requirement. It is not rigorous process simulation and does not
include a full equation of state, fluid-property calculation, or
detailed equipment design.

Engineering Feasibility labels were assigned in a post-hoc qualitative
engineering review of the frozen outputs. The review did not modify the
raw outputs and applied the criteria above: process topology, flow/state
continuity, operation/state consistency, and stated boundary conditions.
This assessment is not a prespecified statistical endpoint.

Within-Method Outcome is PASS, WARN, or FAIL. Here, WARN denotes an
otherwise acceptable result with a non-fatal qualification. It evaluates
successful completion of the method-specific output target together with
engineering acceptability: a DEXPI/XSD-valid representation for Direct
DEXPI, or a structurally valid IR for Engineering IR. No downstream
deterministic IR-to-DEXPI conversion is evaluated in this preprint.

A Direct output receives Within-Method Outcome FAIL when its XML
representation fails the XSD gate, even if the engineering topology
appears acceptable. Conversely, a structurally valid IR output can
receive Within-Method Outcome FAIL when its engineering states are
inconsistent.

\section{6. Results}\label{results}

\subsection{6.1 Case 1: single-pressure
injection}\label{case-1-single-pressure-injection}

The case requires 700,000 standard m³/day of pipeline-quality gas
supplied at 8.0--10.0 MPa and approximately 2 °C to be delivered to
injection wells at up to 35 MPa and no more than 60 °C.

\textbf{Direct DEXPI.} The generated design uses a three-compressor
injection sequence with metering/regulation, two interstage cooling
steps, an aftercooler, and an injection manifold feeding two well sinks:

Metering and Regulation Station → First Stage Compressor → Interstage
Cooler 1 → Second Stage Compressor → Interstage Cooler 2 → Third Stage
Compressor → Aftercooler → Injection Manifold → Injection Wells

The XML is parseable and XSD-valid, but the represented process is
rejected on engineering grounds. The feed connection \texttt{str\_01}
uses \texttt{port\_meter\_in} as both source and target, creating a
self-loop instead of a source-to-meter connection. Most main-process
streams have no flow value, and the two post-manifold well branches also
have no flow values. The aftercooler is represented with a temperature
increase from 48 °C to 55 °C, which contradicts its cooling function.
Although the design contains a third compressor object, the represented
states do not show a clear additional pressure rise through that stage.
Engineering Feasibility is REJECT and the Within-Method Outcome is FAIL
despite successful XSD validation.

\textbf{Engineering IR.} The IR selects a simpler two-stage compression
design:

Regional Transmission Network → Inlet Metering and Filtration → First
Stage Compression → Interstage Cooling → Second Stage Compression →
Aftercooling → Injection Metering and Manifold → Underground Gas Storage
Wells

The JSON is parseable and structurally valid, with 8 nodes and 7
streams. All seven streams carry 700,000 standard m³/day. The first
compressor raises pressure to 18.5 MPa, interstage cooling lowers
temperature from 105 °C to 40 °C, and the second compressor raises
pressure to 35.0 MPa before aftercooling to 50 °C. The final state is
approximately 35 MPa / 50 °C and satisfies the stated boundary. The
remaining concern is a small unexplained pressure sequence after
aftercooling, 35.0 MPa → 34.8 MPa → 35.0 MPa, across metering and
manifolding without an explicit pressure-raising operation for the 34.8
→ 35.0 MPa step. Engineering Feasibility is WARN and the Within-Method
Outcome is WARN.

\subsection{6.2 Case 2: withdrawal and
export}\label{case-2-withdrawal-and-export}

The case requires 450,000 standard m³/day of withdrawn gas supplied at
12.0--14.0 MPa and approximately 25 °C to be exported at 6.5--7.5 MPa
and no more than 45 °C.

\textbf{Direct DEXPI.} The generated design uses inlet isolation and
filtration, followed by pressure letdown and export-gas cooling:

Withdrawal Well Manifold → Inlet ESDV → Inlet Filter Separator →
Pressure Control Valve → Export Gas Cooler → Export Pipeline

The XML is parseable and XSD-valid. A separate \texttt{FT-01} Export
Flow Meter is also generated, but it has no material ports and is not
connected to the main material topology, so it is not part of the actual
connected process path. The connected design reduces pressure from the
withdrawal side to approximately 7 MPa and represents cooling from 48 °C
to 35 °C before export. The final connected stream records 450,000
m³/day at approximately 7 MPa / 35 °C. The numerical flow value matches
the specified capacity, but the standard-volume basis is not explicitly
encoded in that connected stream. Intermediate flow metadata is sparse,
and the pressure-letdown state rises from approximately 25 °C to 48 °C.
Without composition, EOS, or Joule--Thomson data, that rise is treated
as a plausibility concern rather than an unconditional mathematical
contradiction. Engineering Feasibility is WARN and the Within-Method
Outcome is WARN.

\textbf{Engineering IR.} The IR generates a related but not identical
export process:

Withdrawal Well Manifold → Inlet Separator → Pressure Control Valve →
Export Gas Cooler → Export Metering Skid → Regional Transmission
Pipeline

The JSON is parseable and structurally valid, with 6 nodes and 5
streams. In contrast to the Direct output, the export metering step is
part of the connected main path. All five streams carry 450,000 standard
m³/day. The pressure-control valve reduces the represented state from
12.8 MPa / 25 °C to 7.0 MPa / 12 °C, and the final export state is 6.8
MPa / 35 °C, satisfying the stated pressure and temperature boundary.
However, the explicitly named cooling operation raises temperature from
12 °C to 35 °C:

Pressure Control Valve outlet: 7.0 MPa / 12 °C Export Gas Cooler outlet:
6.9 MPa / 35 °C

This direct operation/state contradiction is sufficient for Engineering
Feasibility to be REJECT. The Within-Method Outcome is FAIL.

\subsection{6.3 Case 3: dual-pressure
injection}\label{case-3-dual-pressure-injection}

The case requires simultaneous delivery of 300,000 standard m³/day to
Well Group A at 20 MPa and 600,000 standard m³/day to Well Group B at 30
MPa, with both outlets no warmer than 55 °C; the feed is supplied at
8.0--9.0 MPa and approximately 15 °C.

\textbf{Direct DEXPI.} The generated design first splits the feed and
then uses two parallel compression-and-cooling branches:

Regional Transmission Network → Feed Header Splitter

Group A: Train A Compressor → Train A Aftercooler → Injection Well Group A

Group B: Train B Compressor → Train B Aftercooler → Injection Well Group B

At the equipment-object level, the Train A compressor is specified with
two stages and the Train B compressor with three stages. The XML
therefore represents different compression requirements for the two
pressure levels, although the individual internal stages are not
expanded as separate process nodes. The stream states record 900,000
m³/day splitting into 300,000 and 600,000 m³/day. These numerical values
match the specified design flows and preserve the 300,000 + 600,000 =
900,000 balance, but the standard-volume basis is not explicitly
encoded. Group A ends at 20 MPa / 50 °C and Group B at 30 MPa / 50 °C.
Both branches are present simultaneously, compression raises pressure,
and cooling lowers temperature. Engineering Feasibility is WARN because
the topology and state behavior are acceptable but the required flow
basis is incompletely represented.

The XML is parseable but fails the DEXPI XSD because the validator
reports an unexpected \texttt{Data} element inside an
\texttt{AggregatedDataValue} structure. The Within-Method Outcome is
therefore FAIL.

\textbf{Engineering IR.} The IR expands the compression trains into
individual stages. After source metering and flow splitting, Group A
uses two compression stages and Group B uses three:

Pipeline Gas Supply → Custody Metering Station → Flow Splitter

Group A: Compressor Stage 1 → Intercooler → Compressor Stage 2 → Aftercooler →
Injection Well Group A

Group B: Compressor Stage 1 → Intercooler 1 → Compressor Stage 2 →
Intercooler 2 → Compressor Stage 3 → Aftercooler → Injection Well Group B

The JSON is parseable and structurally valid, with 15 nodes and 14
streams. It represents a 900,000 standard m³/day feed splitting into
300,000 and 600,000 standard m³/day branches. On Group A, pressure
progresses through 8.4, 13.5, 13.2, 20.5, and 20.0 MPa across the split,
compression, cooling, second compression, and final cooling states. On
Group B, the corresponding sequence is 8.4, 14.0, 13.7, 22.0, 21.7,
30.5, and 30.0 MPa. Cooling steps lower the represented temperatures,
and the final states are 20 MPa / 40 °C for Group A and 30 MPa / 40 °C
for Group B. The simultaneous dual-pressure requirement is therefore
represented without an obvious operation/state contradiction.
Engineering Feasibility is ACCEPT and the Within-Method Outcome is PASS.

\subsection{6.4 Unified result table}\label{unified-result-table}

{\def\LTcaptype{none} 
\begin{longtable}[]{@{}
  >{\raggedright\arraybackslash}p{(\linewidth - 8\tabcolsep) * \real{0.2000}}
  >{\raggedright\arraybackslash}p{(\linewidth - 8\tabcolsep) * \real{0.2000}}
  >{\raggedright\arraybackslash}p{(\linewidth - 8\tabcolsep) * \real{0.2000}}
  >{\raggedright\arraybackslash}p{(\linewidth - 8\tabcolsep) * \real{0.2000}}
  >{\raggedright\arraybackslash}p{(\linewidth - 8\tabcolsep) * \real{0.2000}}@{}}
\toprule\noalign{}
\begin{minipage}[b]{\linewidth}\raggedright
Case
\end{minipage} & \begin{minipage}[b]{\linewidth}\raggedright
Method
\end{minipage} & \begin{minipage}[b]{\linewidth}\raggedright
Representation Validity
\end{minipage} & \begin{minipage}[b]{\linewidth}\raggedright
Engineering Feasibility
\end{minipage} & \begin{minipage}[b]{\linewidth}\raggedright
Within-Method Outcome
\end{minipage} \\
\midrule\noalign{}
\endhead
\bottomrule\noalign{}
\endlastfoot
Case 1 & Direct DEXPI & PASS & REJECT & FAIL \\
Case 1 & Engineering IR & PASS & WARN & WARN \\
Case 2 & Direct DEXPI & PASS & WARN & WARN \\
Case 2 & Engineering IR & PASS & REJECT & FAIL \\
Case 3 & Direct DEXPI & FAIL & WARN & FAIL \\
Case 3 & Engineering IR & PASS & ACCEPT & PASS \\
\end{longtable}
}

Direct DEXPI is XSD-valid in two of the three cases. Engineering IR is
structurally valid in all three cases. Engineering outcomes do not
consistently favor either method.

\subsection{6.5 Token and latency
comparison}\label{token-and-latency-comparison}

{\def\LTcaptype{none} 
\begin{longtable}[]{@{}llrrr@{}}
\toprule\noalign{}
Case & Method & Prompt tokens & Completion tokens & Latency (s) \\
\midrule\noalign{}
\endhead
\bottomrule\noalign{}
\endlastfoot
Case 1 & Direct DEXPI & 121,839 & 10,573 & 168.13 \\
Case 1 & Engineering IR & 617 & 590 & 11.45 \\
Case 2 & Direct DEXPI & 121,881 & 6,655 & 87.71 \\
Case 2 & Engineering IR & 659 & 434 & 10.03 \\
Case 3 & Direct DEXPI & 121,921 & 9,467 & 146.78 \\
Case 3 & Engineering IR & 699 & 1,078 & 21.09 \\
\end{longtable}
}

In the evaluated method bundles, deferring DEXPI serialization
substantially reduced prompt burden across all three cases. The
Direct-to-IR prompt-token ratios are approximately 197×, 185×, and 174×
for Cases 1--3, respectively. Latency is reported descriptively and is
not treated as a strict model performance conclusion because server
load, caching, and backend scheduling can affect it.

\section{7. Discussion}\label{discussion}

For the six outputs from two method bundles applied to three cases under
one fixed model snapshot, with one completed generation per condition,
representation validity, engineering feasibility, and method-specific
completion are distinct. Case 1 Direct DEXPI passes the XSD gate but
fails engineering review because of topology and operation/state
defects. Case 3 Direct DEXPI has a broadly coherent dual-pressure
topology but fails XSD validation and incompletely represents the
required standard-volume flow basis. Passing a structural gate therefore
does not establish engineering consistency, and engineering coherence
does not ensure standards-level validity.

The two methods also expose process design at different levels of
detail. Case 1 Direct DEXPI uses three compressor objects with two
interstage coolers, whereas the Engineering IR output uses two
compression stages. In Case 3, Direct DEXPI encapsulates the two- and
three-stage compression requirements inside compressor objects, while
Engineering IR expands individual compression and cooling stages as
separate nodes. Because Direct DEXPI and Engineering IR are method
bundles, these design differences cannot be attributed to the
representation alone.

The clearest observed advantage of the Engineering IR is inspectability
rather than uniformly better engineering correctness. Case 2 Engineering
IR is structurally valid, preserves the full design flow, and reaches
the stated export boundary, yet an explicitly named cooling operation
raises temperature from 12 °C to 35 °C. The compact node-and-stream
representation makes this inconsistency straightforward to localize.
Across the three cases, Engineering IR passes structural validation in
all three while its engineering outcomes range from WARN to REJECT to
ACCEPT.

Representation burden differs sharply between the two method bundles. In
the evaluated method bundles, Direct prompts contain approximately
121.8k--121.9k tokens, compared with 617--699 tokens for Engineering IR,
corresponding to approximately 174×--197× more prompt tokens in the
Direct condition. This difference reflects the method bundles: Direct
DEXPI includes the frozen DEXPI reference context and standards-level
serialization, whereas Engineering IR uses a compact node-and-stream
schema. Deferring DEXPI serialization in these evaluated method bundles
therefore substantially reduces prompt burden and makes serialization
failures easier to distinguish from engineering failures.

\section{8. Limitations}\label{limitations}

This is a diagnostic comparison of one fixed model snapshot on three
test cases, with one completed generation per case-method condition. The
six outputs are therefore evidence about these specific runs rather than
a basis for statistical inference or broad claims about other models and
process systems.

Direct DEXPI and Engineering IR are method bundles rather than a single
controlled-variable comparison. Their prompt context, output contract,
validator, and serialization responsibility differ together; in
particular, the Direct condition includes the large frozen DEXPI
reference context. Consequently, the observed differences cannot be
attributed to XML versus JSON alone.

The engineering review is a post-hoc qualitative assessment of the
frozen outputs, not a prespecified statistical endpoint. It checks
topology, flow and state continuity, operation/state consistency, and
stated boundary conditions, but it is not a full process simulation or
detailed equipment design. No full equation of state or fluid-property
calculation is performed. Likewise, IR structural validity is only a
structural result, and Direct XSD validity does not establish full DEXPI
semantic conformance. No deterministic IR-to-DEXPI converter is
evaluated in this preprint.

\section{9. Conclusion}\label{conclusion}

In this three-case diagnostic comparison of two method bundles, using
one fixed model snapshot and one completed generation per condition,
Direct DEXPI was XSD-valid in two cases, while all three Engineering IR
outputs were structurally valid. The Direct condition exhibited both
engineering inconsistency and standards-level serialization failure,
showing that passing the representation-validity gate and achieving an
acceptable engineering result are separate requirements. Engineering IR
reduced the prompt burden from approximately 121.8k--121.9k tokens to
617--699 tokens and made process objects and state transitions more
compact to inspect, but one IR output still contained an explicit
cooling-state contradiction.

The observed comparison indicates that, within these evaluated method
bundles, separating engineering semantics from DEXPI serialization can
substantially reduce representation burden and make serialization
failures easier to isolate, but it does not by itself eliminate
engineering inconsistencies.

These observations are specific to the completed runs and do not support
statistical or population-level inference.

\section{Data and Code Availability}\label{data-and-code-availability}

The reproducibility package is publicly available at
\url{https://github.com/Qingchuan-ZHU/ugs-dexpi-ir-reproducibility}.

The repository preserves the formal runner, exact prompt templates,
formal case requirements, fixed model configuration, XSD validator,
frozen DEXPI reference artifacts, engineering review, all six raw
generated outputs, their validation records, and the formal SHA-256
manifest. The raw generated outputs are preserved without repair,
normalization, or result-dependent retry. The frozen DEXPI 2.0.0
reference artifacts are redistributed under CC BY 4.0; source
attribution and third-party licensing are recorded in
\texttt{THIRD\_PARTY\_NOTICES.md}.

\texttt{REPRODUCIBILITY.md} documents how to inspect the frozen setup
and how to start a new hosted run. A rerun consumes API calls and is not
expected to reproduce identical output bytes. No real API key is
included in the repository.

\section{Declaration of Generative AI and AI-Assisted
Technologies}\label{declaration-of-generative-ai-and-ai-assisted-technologies}

During the preparation of this work, Qingchuan Zhu used OpenAI ChatGPT
and Codex to assist with manuscript organization, language refinement,
and code and document preparation. Qingchuan Zhu reviewed and edited the
resulting material and takes full responsibility for the content of the
manuscript.

Generative AI models evaluated as part of the study are described
separately in the Methods and are not covered by this
manuscript-preparation disclosure.

\section{Competing Interests}\label{competing-interests}

The authors declare no competing interests.

\section*{References}\label{references}
\addcontentsline{toc}{section}{References}

\protect\phantomsection\label{refs}

\end{document}